\documentclass[aps,prl,twocolumn,showpacs,superscriptaddress,groupedaddress]{revtex4}  % for review and submission
\usepackage{graphicx}  % needed for figures
\usepackage{dcolumn}   % needed for some tables
\usepackage{bm}        % for math
\usepackage{amssymb}   % for math
\usepackage{color}
\begin{document}
%\title{Cooling of binary mixture of Lennard Jones particles}
\title{Cooling-rate dependence of kinetic and mechanical stability of simulated glasses}
\author{Hannah Staley}
\affiliation{Department of Physics, Colorado State University, Fort Collins, Colorado 80523, USA}
\author{Elijah Flenner}
\author{Grzegorz Szamel}
\affiliation{Department of Chemistry, Colorado State University, Fort Collins, Colorado 80523, USA}

\date{\today}

\begin{abstract}
Recently, ultrastable glasses have been created through vapor deposition. Subsequently, computer simulation algorithms have been proposed that mimic the vapor deposition process and result in simulated glasses with increased stability. In addition, random pinning has been used to generate very stable glassy configurations without the need for lengthy annealing or special algorithms inspired by vapor deposition. Kinetic and mechanical stability of experimental ultrastable glasses is compared to those of experimental glasses formed by cooling. We provide the basis for a similar comparison for simulated stable glasses: we analyze the kinetic and mechanical stability of simulated glasses formed by cooling at a constant rate by examining the transformation time to a liquid upon rapid re-heating, the inherent structure energies, and the shear modulus. The kinetic and structural stability increases slowly with decreasing cooling rate. The methods outlined here can be used to assess kinetic and mechanical stability of simulated glasses generated by using specialized algorithms.
\end{abstract}

\pacs{64.70.Q-,81.05.Kf}
\maketitle

%\section{Introduction}
Vapor deposition on a substrate held at around 85\% of the glass transition temperature is used to create glasses that have a higher kinetic stability \cite{E_S} and larger elastic moduli \cite{E_AM} than glasses created by cooling from a liquid. Due to these and other desirable properties of these so-called ultrastable glasses, it is of great interest to understand the differences between glasses created through vapor deposition and by cooling from a liquid. Since vapor deposited glasses are created one layer at a time, it is suspected that enhanced mobility at the surface allows the molecules to find more stable configurations. 

Discovery of ultrastable glasses in the lab inspired several computer simulation studies. The goal of these studies has been two-fold. First, modeling the vapor deposition process can provide insight into the origin of the stability of vapor deposited glasses. Second, the experimental procedure can be used as an inspiration for the development of computer simulation algorithms that could generate very stable simulated glasses without the need for lengthy annealing. L\'{e}onard and Harrowell \cite{LH} modeled ultrastable glasses using a three-spin facilitated Ising model. Their study provided evidence that enhanced surface mobility during vapor deposition results in improved kinetic stability, but the simplicity of their model prevented them from studying other properties of ultrastable glasses, such as the shear modulus. On the other end of the complexity spectrum, Singh \textit{et al.} \cite{SdP} modeled vapor deposition of the organic molecule trehalose by introducing 1-5 ``hot'' molecules above a substrate and then (artificially) cooling them to the substrate temperature. This procedure resulted in a simulated glass with a higher kinetic stability, but also pronounced structural anisotropy in the direction perpendicular to the substrate.

More recent studies used well-known glass-forming binary Lennard-Jones mixtures \cite{ka}. First, Singh, Ediger, and de Pablo \cite{E_NM} and Lyubimov, Ediger, and de Pablo \cite{E_JCP} modeled the vapor deposition of a binary mixture of Lennard-Jones particles to simulate the creation of an ultrastable glass. They found an enhanced kinetic stability compared to the glass obtained by cooling from a liquid at a constant rate, but the enhanced stability was less pronounced than in experiments. Furthermore, the simulated vapor deposition resulted in glasses with structural anisotropy, a higher density, and non-uniform composition compared to the simulated glasses created by cooling from a liquid. Very recently, Hocky \textit{et al.} \cite{H_X} created two dimensional binary Lennard-Jones mixture glasses by randomly pinning particles. Glasses formed by random pinning are, by construction, isotropic and uniform (on average). Hocky \textit{et al.} found that these glasses have increased kinetic stability but the presence of pinned particles prevented them from investigating whether these glasses also have a larger shear modulus.

The above mentioned simulational studies focused on the properties of stable glasses created by using different specialized algorithms. However, it is difficult to assess the stability of these glasses without knowledge of the stability of simulated glasses generated in a conventional way, \textit{i.e.} by cooling at a constant rate. We emphasize that simulated glasses generated by cooling are isotropic and uniform. Thus, studying these glasses may also shed some light on the importance of the anisotropy and compositional non-uniformity for the glass stability. Here we create glasses by cooling from the supercooled liquid and we assess their kinetic and mechanical stability by examining their kinetic stability against reheating, properties of their potential energy landscape, and their shear modulus.

%\section{Methods}

We simulated the Kob-Andersen (KA) \cite{ka} 80:20 binary Lennard-Jones mixture in three dimensions. The particles interact via the potential 
%\begin{equation}
\label{eq:ljpot}
$V_{\alpha\beta}(r) = 4\epsilon_{\alpha\beta} 
\left[\left(\frac{\sigma_{\alpha\beta}}{r}\right)^{12} - \left(\frac{\sigma_{\alpha\beta}}{r}\right)^6\right]$, 
%\end{equation}
and the parameters are $\epsilon_{BB} = 0.5 \epsilon_{AA}$, $\epsilon_{AB} = 1.5 \epsilon_{AA}$, $\sigma_{BB} = 0.88 \sigma_{AA}$, and $\sigma_{AB} = 0.8 \sigma_{AA}$. The majority species are of type $A$ and both masses are equal. We cut the potential at $2.5\sigma_{\alpha\beta}$. The results are presented in reduced units with $\sigma_{AA}$, $\epsilon_{AA}/k_B$, and $\sqrt{m_A\sigma_{AA}^2/\epsilon_{AA}}$ being the units for length, temperature, and time, respectively. We simulated $N=8000$ particles at a number density $N/V = 1.20$ using periodic boundary conditions. 

We ran NVT simulations using a Nos\'{e}-Hoover thermostat. We started with an equilibrated supercooled liquid at the temperature $T = 0.5$ (for this system the onset temperature for the slow dynamics is $T_o \approx 1.0$ and the mode-coupling temperature is $T_c=0.435$). We ran 80 cooling runs at cooling rates $\dot{T} = \Delta T/\Delta t$ of $3.33 \times 10^{-n}$ where $n=3$, 4, 5, 6, and 7. Each run started at an equilibrium configuration at $T=0.5$ and was cooled to $T= 0.3$. We also ran 4 cooling runs at a cooling rate of $3.33 \times 10^{-8}$. For $\dot{T} > 3.33 \times 10^{-8}$, we started from the final configuration of each of the 80 cooling runs and annealed the system for at least 100 times $\tau_{\alpha}$ determined from $T = 0.5$ (we define $\tau_\alpha$ through the average overlap function discussed later). For the cooling rate of  $3.33 \times 10^{-8}$, we performed 5 different annealing runs for each of the 4 configurations, each beginning with a different initial random velocity. In addition, we heated the final configurations of the cooling runs by ramping up the temperature from $T = 0.3$ to $T = 0.5$ at a constant rate over a time $t = 10$ (we tried increasing the temperature instantaneously from $T = 0.3$ to $T = 0.5$ but found that this resulted in large, non-physical, oscillations of the kinetic and potential energies). After ramping up the temperature to $T=0.5$, we continued constant temperature simulations (at $T=0.5$) until the mean-square displacement started growing linearly with time. We refer to this ramping up of temperature and the subsequent run at $T = 0.5$ as a heating trajectory. For $\dot{T} > 3.33 \times 10^{-8}$, we ran 80 heating trajectories, in each case starting from the final configuration of a different cooling run. For $\dot{T} = 3.33 \times 10^{-8}$, we ran the heating trajectories 15 times using different initial random velocities for each of the configurations obtained from the 4 cooling runs. Finally, we also performed an equilibrium run at $T = 0.5$, for at least $100\tau_{\alpha}$. We used HOOMD-blue \cite{h_o, h_a} for equilibration at $T=0.5$ and the cooling simulations. The heating trajectories, the $T = 0.3$ runs, and the single run at $T = 0.5$ used LAMMPS \cite{L_o, L_a} run on a GPU \cite{B0, B11, B12}.

%\section{Results}
We start with an assessment of the kinetic stability of the simulated glasses. First, we we examine the time it takes for the particles to rearrange after the sudden increase in temperature, \textit{i.e.} along the heating trajectories. We quantify these rearrangements through the average overlap function $q_{s} (t,t_{w})$, which measures the probability that a particle moved over distance $a$ during the time between $t_w$ and $t+t_w$, where $t_w$ is the waiting time.
\begin{equation}
\label{eq:qs}\label{qs}
q_{s} (t,t_{w}) = \frac{1}{N} \left\langle \sum_{m} q_{m} (t,t_{w}) \right\rangle ,
\end{equation}
where $q_{m} (t,t_{w}) = \Theta \left( a - \left| \mathbf{r}_{m} (t + t_{w}) - \mathbf{r}_{m} (t_{w}) \right| \right)$, $\Theta$ is Heaviside's step function, and $\mathbf{r}_{m}(t)$ is the position of a particle $m$ at a time $t$. $t_{w}$, the waiting time, is measured from the start of the heating trajectory. We chose $a = 0.25$ to be consistent with previous work \cite{FS}. We define $\tau_{\alpha}$ as when $q_{s} (\tau_{\alpha},t_w) = e^{-1}$ for an equilibrium system (note that for an equilibrium system $q_{s} (t,t_{w})$ does not depend on the waiting time, \textit{i.e.} it is time-translationally invariant).

\begin{figure}
\includegraphics[scale=0.3]{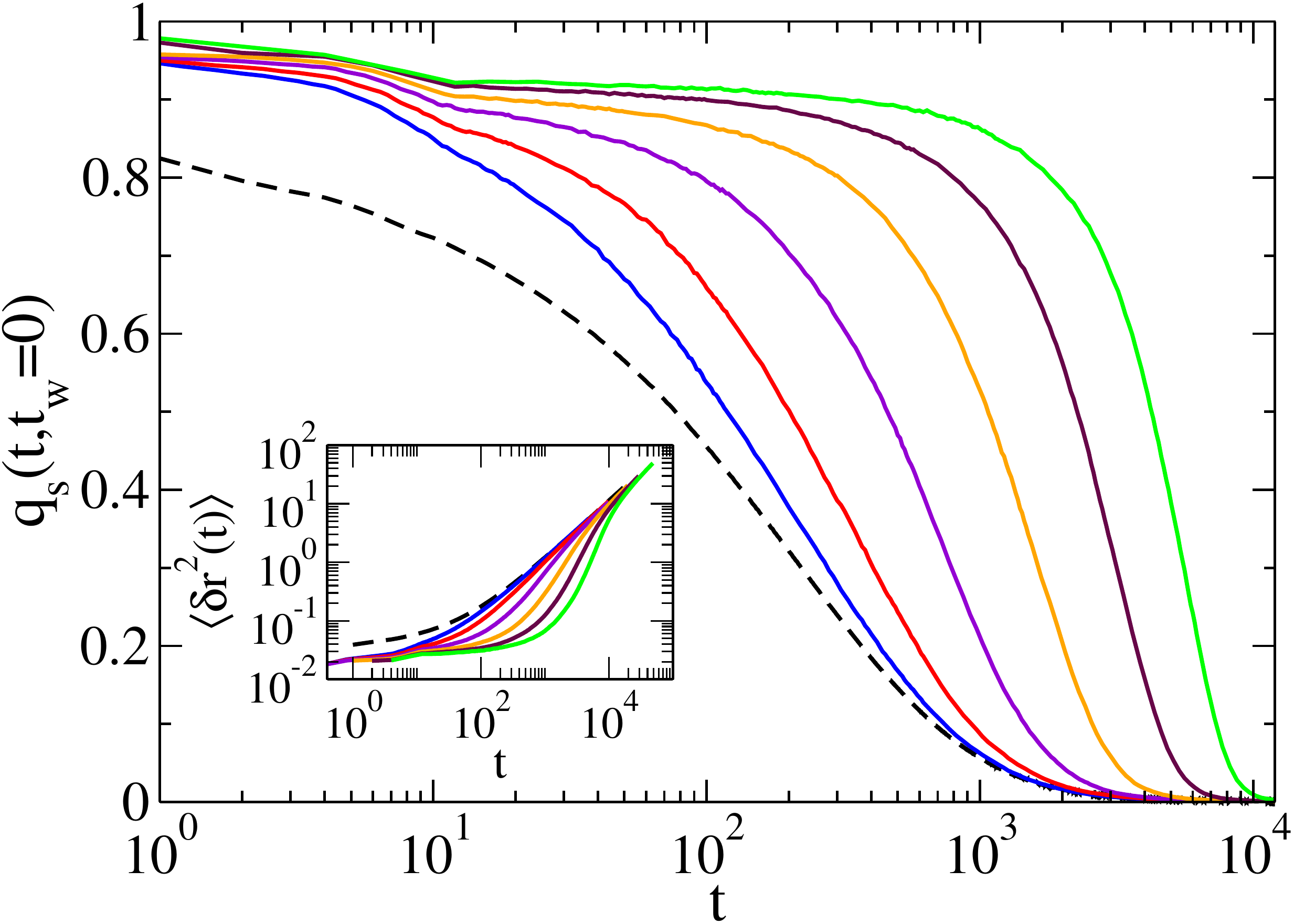}
\caption{\label{fig:theta} Average overlap function $q_s(t,t_w=0)$ monitored during the heating trajectories (solid lines) compared to the average overlap function for the equilibrium liquid at $T=0.5$ (dashed line). Solid lines correspond to glasses obtained with cooling rates  $\dot{T} = 3.33 \times 10^{-n}$, where $n = 3$, 4, 5, 6, 7, and 8 from left to right. The inset shows the mean square displacement versus time for the same cooling rates (solid lines) and for the equilibrium liquid at $T=0.5$ (dashed line).}
\end{figure}

Fig.~\ref{fig:theta} shows average overlap function $q_s(t,t_w=0)$ obtained while heating glasses prepared at different cooling rates (solid lines) and obtained from the equilibrium run at $T = 0.5$ (dashed line). The feature at $t=10$ is a consequence of changing the thermostat from a ramp-up of the temperature at a constant rate from $T=0.3$ to $T=0.5$ to maintaining constant temperature $T=0.5$. For $t>10$, $q_s(t,t_w=0)$ exhibits prolonged plateaus for the three slowest cooling rates followed by a very rapid decay at later times (the glass generated using the slowest cooling rate,$3.33 \times 10^{-8}$, relaxes faster than for ballistic motion). With decreasing cooling rate both the plateau height and its extent increase. This indicates that with decreasing cooling rate the cage diameter decreases and it takes longer for the particles to move out of their cages. This is the first indication of increasing kinetic stability against re-heating. 

The inset in Fig.~\ref{fig:theta} shows the mean square displacement $\left< \delta r^{2} (t,t_w) \right\rangle = N^{-1} \left\langle \sum_{n} \left[ \mathbf{r}_{n} (t+t_w) - \mathbf{r}_{n} (t_w) \right]^2 \right>$ for a waiting time $t_w = 0$  obtained while heating glasses prepared at different cooling rates (solid lines) and obtained from the equilibrium run at $T = 0.5$ (dashed line). We find that $\left< \delta r^2(t,t_w) \right>$ encodes similar information as $q_s(t,t_w=0)$. In particular, we note that with decreasing cooling rate the time required for the mean-square displacement to reach the equilibrium curve increases. This is another indication of increasing kinetic stability against re-heating.

\begin{figure}
\includegraphics[scale=0.3]{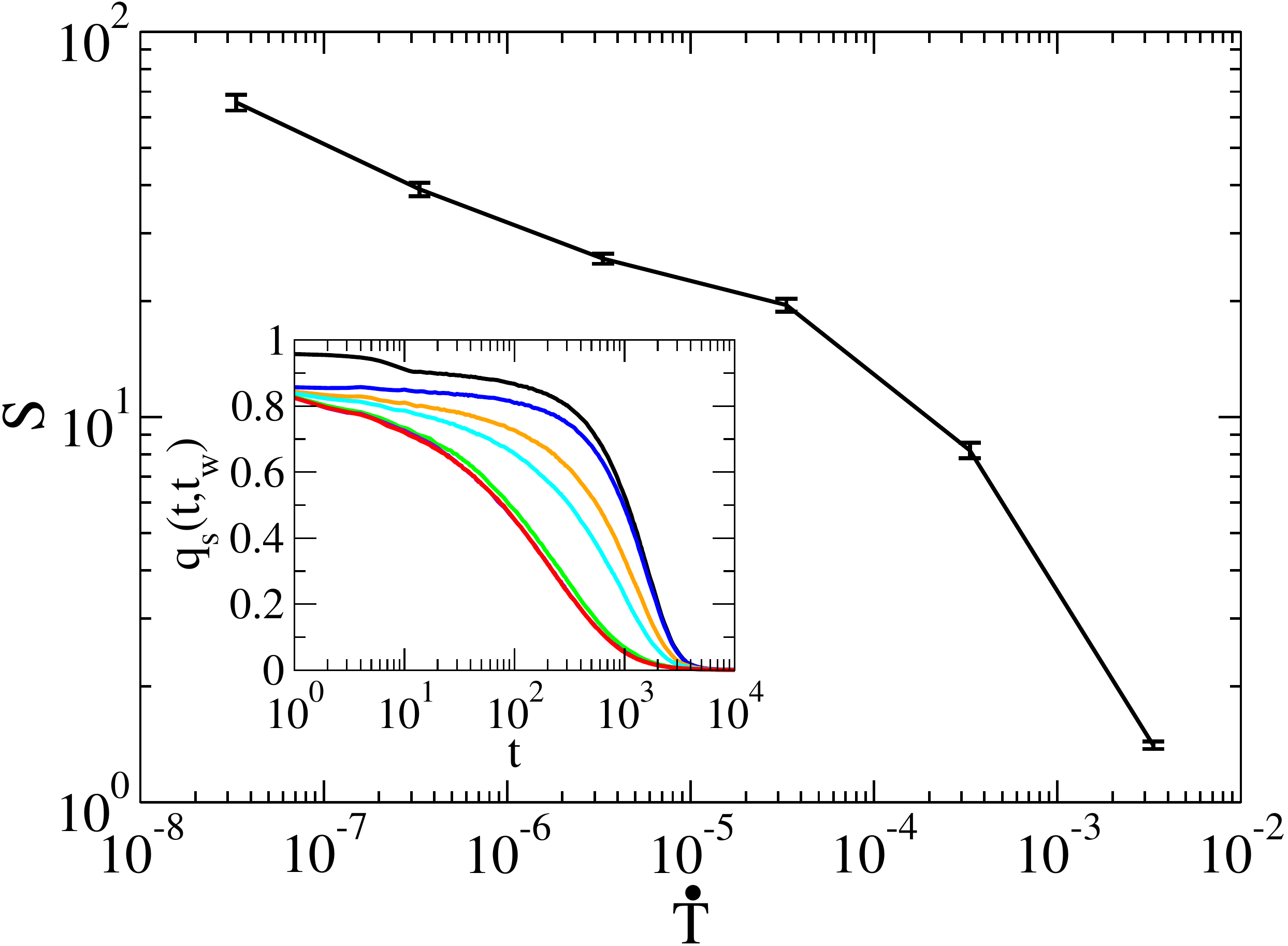}
\caption{\label{fig:S}Dependence of stability ratio $S = t_{trans} / \tau_{\alpha}$ on cooling rate $\dot{T}$. The inset shows the average overlap function versus time for the cooling rate $\dot{T}=3.33 \times 10^{-6}$ and waiting times, from top to bottom, 0, 10, 500, 1000, 3000, 4250, 5000.}
\end{figure}

To quantify kinetic stability we follow earlier experimental \cite{Sep} and simulational \cite{H_X} studies and we evaluate the transformation time $t_{trans}$ and the stability ratio $S$. The transformation time is defined as the time it takes for the liquid obtained by heating the glass to return to equilibrium. In the experimental study of Ref. \cite{Sep} the transformation time was obtained by monitoring the time evolution of the dielectric response. We follow the simulational study of Ref. \cite{H_X} and obtain the transformation time from monitoring the waiting time dependence of the average overlap function. Specifically, we calculate $q_{s} (t,t_{w})$ for a range of $t_w$ (see the inset to Fig.~\ref{fig:S} for $q_{s} (t,t_{w})$ for various waiting times for $\dot{T}=3.33 \times 10^{-6}$) and we define the waiting time-dependent relaxation time $\tau_s(t_w)$ through the equation $q_s(\tau_s,t_w) = e^{-1}$. We define the transformation time, $t_{trans}$, as the waiting time such that $\tau_s(t_w) = \tau_{\alpha}$, where $\tau_{\alpha}$ is the equilibrium relaxation time. Finally, we follow Refs. \cite{H_X,Sep} and we define a stability ratio $S = t_{trans}/\tau_\alpha$.

In Fig.~\ref{fig:S} we show the cooling rate dependence of the stability ratio. At larger cooling rates $S$ increases rather quickly with decreasing $\dot{T}$. However, for cooling rates less than $3.33\times10^{-5}$ we get a slower increase of the stability ratio with cooling rate. For the slowest cooling rates, an order of magnitude decrease in the cooling rate results in a doubling of the stability ratio. Thus, to achieve stability ratios of the most stable simulated glasses, which approach approximately 400 \cite{H_X}, we would need to decrease the cooling rate by about 3 decades (to achieve stability ratios of experimental ultrastable glasses, which approach approximately $10^{3.5}$ \cite{Sep} we would need to decrease the cooling rate by about 22 decades). We conclude that even simulated stable glasses would be impossible to obtain by cooling, at least with    the computational resources available to us at present.

\begin{figure}
\includegraphics[scale=0.3]{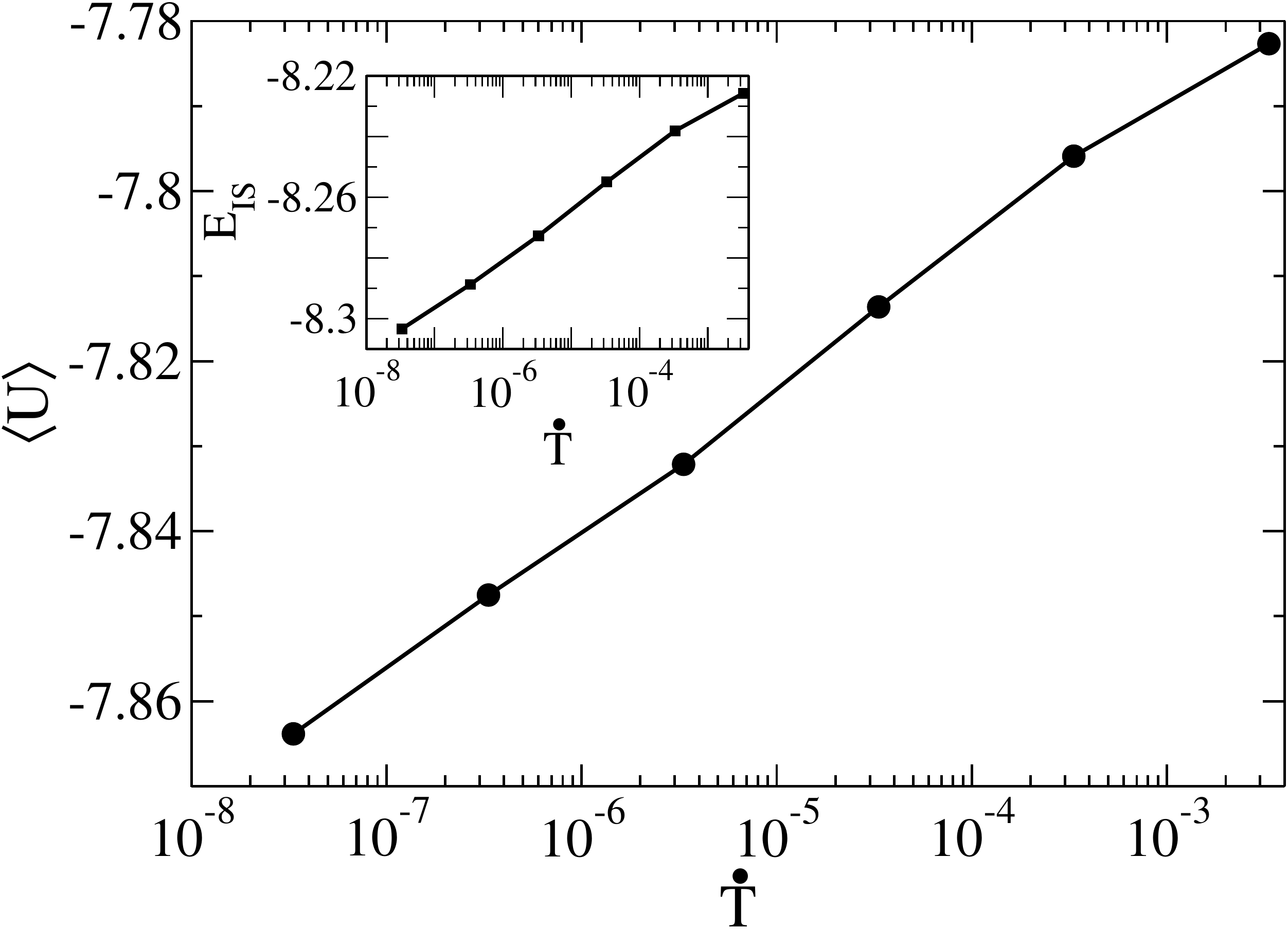}
\caption{\label{fig:U}Average potential per particle, $\langle U \rangle$, as a function of cooling rate, $\dot{T}$. The inset shows average inherent structure energy per particle, $E_{IS}$, as a function of cooling rate. The calculations were done at cooling rates $3.33 \times 10^{-n}$, with $n=8,7,6,5,4,3$, the lines are to guide the eye.}
\end{figure}

Next, we turn to an assessment of the mechanical stability of the simulated glasses. The simplest, albeit indirect, measure of the mechanical stability is provided by the inherent structure energy \cite{Scio}, since in the energy landscape picture of the glass transition, more stable glasses are in a lower energy basin of the landscape and have to overcome a larger energy barrier to deform. We show the dependence of the average inherent structure energy (obtained using the Fire algorithm \cite{fi} in HOOMD-blue) at $T=0.3$ on the cooling rate used to reach this temperature in the inset in Fig.~\ref{fig:U}. We see that the average inherent structure energy decreases with decreasing cooling rate. In  Fig.~\ref{fig:U} we show the dependence of the average potential energy per particle at $T=0.3$ on the cooling rate. It also decreases with decreasing cooling rate. We note that for their vapor deposited glass at $T = 0.3$, Ediger and de Pablo got an inherent structure energy of -8.35 \cite{E_JCP} and a potential energy of -7.8 (estimated from \cite{E_JCP}). These values are close to our values of -8.30 and -7.86. We note, however, that the vapor deposition inspired algorithm used in Ref. \cite{E_JCP} resulted in glasses that were non-uniform in contrast to our simulated glasses, and thus, a literal comparison of the two studies is impossible.

\begin{figure}
\includegraphics[scale=0.3]{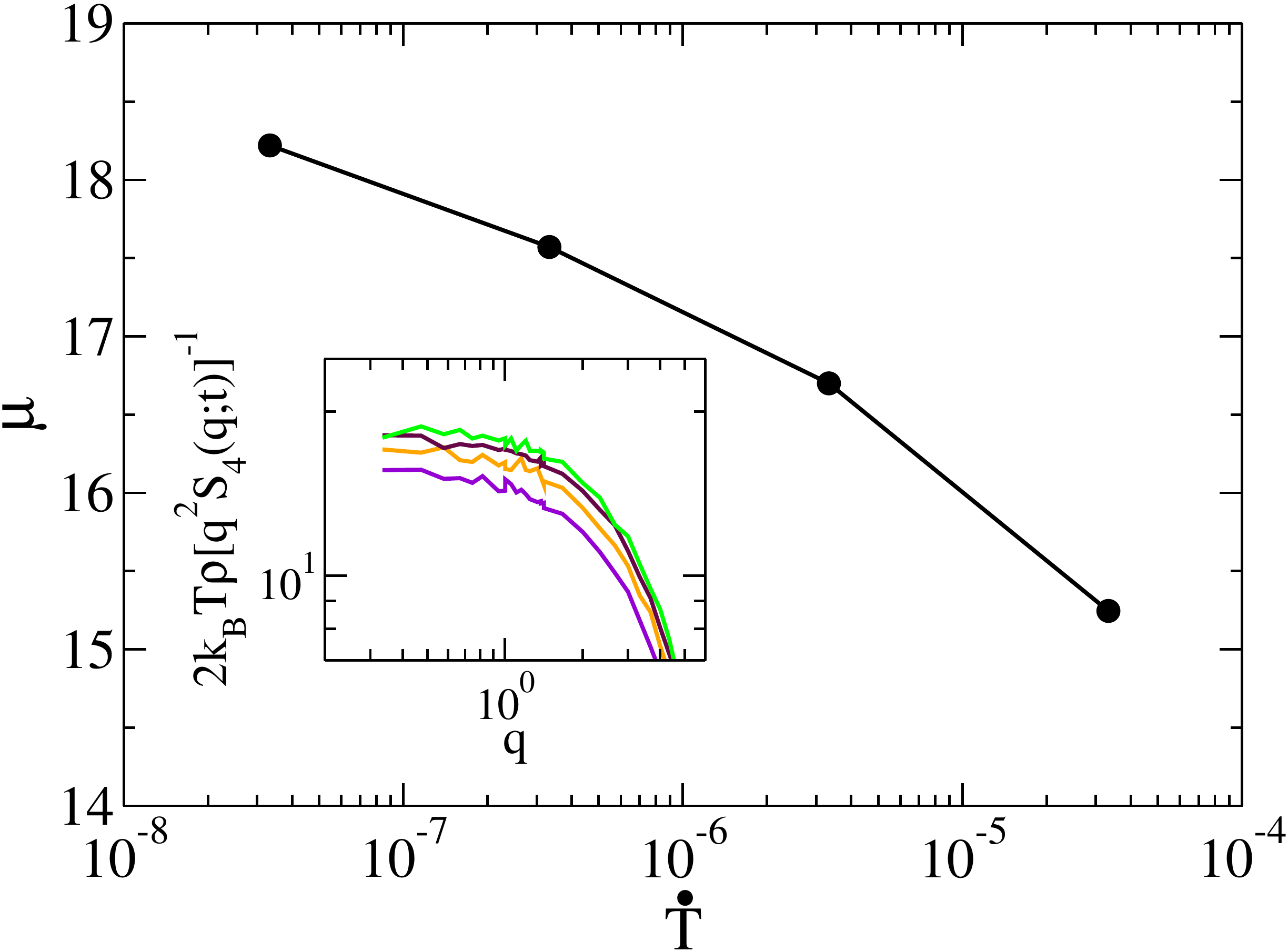}
\caption{\label{fig:mu}Dependence of the shear modulus, $\mu$, on the cooling rate $\dot{T}$ for glasses at $T=0.3$. The inset shows $2 k_{B} T \rho \left[ q^{2} S_{4} (q;t) \right]^{-1}$ versus $q$ plotted at a time $t=100$ for $T = 0.3$, for the four slowest cooling rates: $3.33 \times 10^{-n}$ where $n=5,6,7,8$, from bottom to top.}
\end{figure}

A more direct measure of mechanical stability is provided by elastic constants. Ultrastable glasses prepared in experiments \cite{E_AM} and stable glasses obtained in simulations using vapor deposition-inspired algorithms \cite{E_NM} have larger elastic constants than glasses prepared by ordinary cooling. We examine here the cooling rate dependence of the shear modulus of our simulated glasses. To this end we use an approach recently introduced by two of us \cite{F_X}. This method is based on the relationship between correlations of transverse particle displacements in the glass and the glass's shear modulus. The final result is that the shear modulus of a glass can be measured by studying the small wave-vector $q$ behavior at long times of the four-point structure factor
\begin{equation}
\label{eq:S4}
S_{4} (\mathbf{q};t) = \frac{1}{N} \left\langle \sum_{n, m} g \left[ \delta \mathbf{r}_{n} (t) \right] 
g^{\ast} \left[ \delta \mathbf{r}_{m} (t) \right] 
e^{i \mathbf{q} \cdot \left[ \mathbf{r}_{n} (0) - \mathbf{r}_{m} (0) \right]} \right\rangle ,
\end{equation}
where the weighting function
%\begin{equation}
%\label{eq:g}
$g \left[ \delta \mathbf{r}_{n} (t) \right] = \delta \mathbf{r}_n^{\perp}(t) 
= \left(\mathbf{r}_{n}(t) - \mathbf{r}_{n}(0)\right)^{\perp}$
%\end{equation}
and $\delta \mathbf{r}_n^{\perp}(t)$ is the component of the displacement of particle $n$ perpendicular to the wave-vector $\mathbf{q}$, $\delta \mathbf{r}_n^{\perp}(t) \cdot \mathbf{q} = 0$. As argued in Ref.~\cite{F_X}, the shear modulus for a glass can be obtained from the small $q$, long $t$ behavior of $S_{4} (\mathbf{q};t)$, $\mu = \lim_{q \rightarrow 0} \lim_{t \rightarrow \infty} 2 k_{B} T \rho \left[ q^{2} S_{4} (q;t) \right]^{-1}$. In practice one can examine the small wave-vector behavior of $2 k_{B} T \rho \left[ q^{2} S_{4} (q;t) \right]^{-1}$ for times longer than the initial transient dynamics but shorter than time scale for the aging process of a glass. Also, in order to get good statistics one either needs many more independent configurations at the end of each cooling process than the 80 that we generated or one needs to continue running at $T=0.3$ and average over different time origins. In order to measure properties of the glass as prepared using a specific cooling rate, one can only average over different time origins before the start of the aging process. For faster cooling rates the aging starts very quickly and the range of available time origins is rather small. Fortunately, the present method to calculate the shear modulus does not require very long trajectories \cite{F_X}. We show $2 k_{B} T \rho \left[ q^{2} S_{4} (q;t) \right]^{-1}$ versus $q$ at a time $t = 100$ for $T=0.3$ in the inset to Fig.~\ref{fig:mu}. We averaged over time origins for $t \le 100$ for the 3 faster cooling rates, and for $t \le 400$ for the slowest cooling rate. 

In Fig.~\ref{fig:mu} we show the cooling rate dependence of the shear modulus. We find that the modulus increases with decreasing cooling rate, although the rate of increase seems to be decreasing with decreasing $\dot{T}$. At the slowest cooling rates an order of magnitude decrease of the cooling rate results in an increase of the modulus by 4.6\%. We note that Ashwin, Bouchbinder, and Procaccia \cite{ABP} also investigated the cooling rate dependence of the shear modulus. They used a different system, in two spatial dimensions and thus their results cannot be literally compared to ours. However, we do note that in a similar range of cooling rates our shear modulus increases by 15\% and theirs increases by 12\%. Interestingly, these relative increases of the shear modulus achieved in our and Ashwin \textit{et al.} simulational studies are comparable to the relative differences between the shear modulus of experimental ultrastable glasses and experiemntal glasses generated by cooling (19\% for indomethacin and 15\% for tris-naphtylbenzene \cite{E_AM}). It is unclear to us whether this fact has some deeper meaning. 

%\section{Summary}

In summary, we showed here that by decreasing the cooling rate in simulations, we generated simulated glasses that have increasing kinetic and mechanical stability. The highest kinetic stability that we achieved is about 66. Decreasing the cooling rate by one decade, which would result in the kinetic stability of about 130, would require approximately 1 year on a single GPU using HOOMD-blue per one cooling trajectory. This result establishes a baseline against which specialized algorithms that are inspired by vapor deposition or use random pinning should be measured. 

We believe that the kinetic stability, quantified by the stability ratio $S=t_{trans}/\tau_\alpha$ is the best quantity to compare stability of glasses generated using different protocols and algorithms. We note that the average potential energy is very sensitive to the average density, compositional nonuniformities and possibly to anisotropy. In fact, our glasses generated by cooling at constant rate have lower average potential energy than stable glasses generated by Singh \textit{et al.} \cite{E_NM} using a vapor deposition inspired algorithm. We note the average inherent structure energy seems to correlate better with the stability but most likely it is also sensitive to the density and compositional nonuniformities. Finally, we note that the relative change of the shear modulus of simulated glasses formed by cooling at different cooling rates is comparable to the relative difference between the shear modulus of experimental glasses created through vapor deposition and by cooling. This surprising finding, which deserves further investigation, suggests that that the kinetic and mechanical stability are not necessarily correlated.

The methods presented here can be used to assess kinetic and mechanical stability of simulated glasses generated by using specialized algorithms. 
 
We gratefully acknowledge the support of NSF grant CHE 1213401. 
%This research utilized the CSU IS- TeC Cray HPC System supported by NSF Grant CNS- 0923386.

\bibliography{kabib}
\bibliographystyle{unsrt}

\end{document}